\documentclass{elsart}
\usepackage{graphicx}

\begin{document}
\def\ltsimeq{\,\raise 0.3 ex\hbox{$ < $}\kern -0.75 em
 \lower 0.7 ex\hbox{$\sim$}\,}
 \def\gtsimeq{\,\raise 0.3 ex\hbox{$ > $}\kern -0.75 em
  \lower 0.7 ex\hbox{$\sim$}\,}

\def\avg #1{\langle #1\rangle}

\begin{frontmatter}
\title{Intracultural diversity in a model of social dynamics}
\author{A. Parravano, H. Rivera-Ramirez, and M. G. Cosenza}
\address{Centro de F\'{\i}sica Fundamental,
Facultad de Ciencias, Universidad de Los Andes, \\ Apartado Postal 26
La~Hechicera, M\'erida, M\'erida~5251, Venezuela,}

\begin{abstract}
We study the consequences of introducing individual nonconformity in social interactions,
based on Axelrod's model for the dissemination of culture.
A constraint on the number of situations in which
interaction may take place is introduced in order to lift the unavoidable homogeneity
present in the final configurations arising in Axelrod's related models. 
The inclusion of this constraint leads to the occurrence of
complex patterns of intracultural diversity whose
statistical properties and spatial distribution are
characterized by means of the concepts of cultural affinity and cultural cline.
It is found that the relevant quantity that determines the properties of
intracultural diversity is given by the fraction of cultural features that
characterizes the cultural 
nonconformity of individuals.

\end{abstract}

\end{frontmatter}

\section{Introduction.}

Recently, many dynamical models, inspired by
analogies with physical systems, have been proposed to describe a
variety of phenomena occurring in social dynamics
\cite{Weidlich,Stauffer,Arrow,MaxiR}. 
Examples include the emergence of cooperation and
self-organization, propagation of information and epidemics, opinion
formation, economic exchanges and evolution of social structures. 
In this context,
Axelrod's model \cite{axel97} for the dissemination of culture 
among interacting agents in a social system has attracted much attention
among physicists.

The concept of culture introduced by Axelrod refers to a set of individual 
features or attributes that are subject to social influence.
Agents can interact with their neighbors in the system according to the
cultural similarities that they share. 
From the point of view of statistical physics, this model is appealing 
because it exhibits a nonequilibrium transition between an ordered final 
frozen state (a global homogeneous culture) and a disordered (culturally fragmented) one 
\cite{Castellano00,Vilone,Klemm03a,Maxi3}.
Several extensions of this model have recently been investigated.
For example, cultural drift has been modeled as noise acting on the
frozen disordered configurations \cite{Klemm03b}.
The effects of mass media has been considered as external \cite{GCT}
or autonomous \cite{Shibanai,PRE} influences acting on the system.
The role of the topology of the social network of interactions
have also been addressed \cite{Klemm03a,Greig,Kuperman}. 
Other extensions include the consideration of
quantitative instead of qualitative values for the cultural traits
\cite{Flache1}.
These studies have revealed that Axelrod's model is robust in the sense
that its main properties persists in all those cases.
In particular, the final frozen states invariably consist of one or more
homogeneous cultural groups. 

In this paper, we introduce a constraint on the number of situations in which
interaction may take place, in order to lift the unavoidable homogeneity
in the final states of the above models. Our model is 
motivated by the idea that generally
individuals tend to maintain a minimum degree of identity by keeping
some cultural features different from those of their neighbors. 
This restriction naturally leads to the persistence of complex patterns of diversity
in the cultural groups present in the final state of the system.   

The model is explained in Section 2.
In Section 3, the results of numerical simulations are presented, showing
the patterns of diversity in the final
frozen states. The statistical properties that characterize intracultural
diversity are calculated in Section 4.
Conclusions are presented in Section 5.

\section{Axelrod's model with intracultural diversity.}

Axelrod's model \cite{axel97} considers a square lattice
network of $N=L^2$ elements with open boundaries and nearest neighbor interactions.
The state of element $i$ is given by a cultural vector of $F$ 
components (cultural features)
$(s_{i1},s_{i2},\cdots,s_{iF})$. Each component $s_{if}$
can adopt any of $q$ integer values (cultural traits) in the set $\{1,\dots,q\}$.
Starting from a random initial state
the network evolves at each time step following these simple rules:
i) An element $i$ and one of its four neighbors $j$ is selected at random. 
ii) If the overlap, defined as $\omega(i,j) = \sum_{f=1}^F \delta_{s_{if},s_{jf}}$,
(number of shared features) is in the range $0<\omega(i,j)<F$,
the pair $(i,j)$ is said to be active with a probability of interaction equal
to $\omega(i,j)/F$.
iii) In case of interaction, one of the unshared features $k$ is selected 
at random and element $i$ adopts the trait $s_{jk}$, 
thus decreasing in one unit the overlap of the pair $(i,j)$.

In any finite network the dynamics settles into a
frozen state, characterized by either $\omega(i,j)=0$ or $\omega(i,j)=F$,
$\forall i,j$. Homogeneous or monocultural states correspond to
$\omega(i,j)=F$,  $\forall i,j$, and obviously there are $q^F$ possible
configurations of this state.  Inhomogeneous or multicultural
states consist of two or more homogeneous domains interconnected
by elements with zero overlap. A domain, or a cultural region, 
is a set of contiguous sites with identical cultural traits.
Castellano et al. \cite{Castellano00} 
demonstrated that the final states of the system experience a  
transition from ordered states (homogeneous culture) for $q < q_c$ 
to disordered states (cultural fragmented)  for $q > q_c$, where $q_c$
is a critical value that depends on $F$.

In order to allow for diversity we introduce a 
parameter $F_d$ such that
a pair $(i,j)$ is considered active if the overlap is in the 
range $0<\omega(i,j)<F-F_d$,
with $0<F_d<F$. 
There is no restriction on which of the $F_d$ features cannot be exchanged
by an active pair. 
The case $F_d=0$ recovers the original Axelrod's model, whereas
$F_d=F$ results in frozen configurations for all possible initial
configurations.
The number of possible frozen states is the number of configurations
in which neighbors cannot longer interact; thus increasing $F_d$ results in an
increase of this number.
The parameter $F_d$ reduces the number of situations in which interactions
may take place. 
In the context of social dynamics,
the ratio $F_d/F$ can be seen as a measure of individual nonconformity.

A cultural region is a set of contiguous sites that possess the same cultural 
vector, whereas a cultural zone is defined as a set of contiguous
sites that share one or more cultural traits; elements in a cultural zone
are said to have a ``compatible" culture \cite{axel97}.
Cultural zones appear as transient states in the original Axelrod's model,
but in the final state only cultural regions are present.   
When $F_d > 0$, cultural zones will usually be present in the final
state because then contiguous sites have an overlap $\omega(i,j)\geq F-F_d$.
The model can be modified by fixing in advance a subset of features that the elements 
of a cultural zone must share in the final state.
  
\section{Numerical Results}

As an example of the effects resulting from the inclusion of the parameter $F_d$
in Axelrod's model, we shall consider a system of size
$N=20 \times 20$ with $F=11$, and $q=10$, starting from random initial 
conditions.
For $F_d=0$, the system converges to a homogeneous state, i.e., 
a single cultural region, since $q << q_c \sim 60$.  
For $F_d=1$ the final state consists of a single cultural zone possessing  
few surviving traits. 
We denote by $K_{f}(s)$ the number of times that
the trait value $s$ appears in the $f$th feature of the cultural vectors in the system.
That is, $K_{f}(s)= \sum_{i=1}^N \delta_{s_{if},s}$.
For a particular realization in a system of size $N=20 \times 20$,
Table 1 shows $K_{f}(s)$ for the final state.
The first row in Table 1 shows that $K_{1}(1)=N$ and for the
remaining traits $K_{f=1}(s\neq 1)=0$; that is, 
all the $400$ cultural vectors have reached the value $1$ in their first feature.
Note that all $400$
elements also share the traits associated to features $f=6,9$, and $10$.
In features  $f=3,4,7,8$, and $11$ only $2$ different values of traits survive; while in
features  $f=2$ and $5$ there appear $3$ different values of traits. 
The fraction of trait values that disappear during the evolution 
towards the final state of the system is $\sum_{f,s}\delta_{K_{f}(s),0}/(f \times (q+1))$.
Therefore, for the realization in Table 1, about $82\%$
of all the trait values existing in the cultural vectors at the beginning
have disappeared in the final state of the system.

In the final state there are 47 different cultural vectors in the system. 
Unexpectedly, the abundance of these vectors follows a
nonuniform distribution.
Table 2 shows the distribution of 
the 15 most abundant cultural vectors in the final state. These vectors 
are ranked according
to their abundance. The abundance of the vector of rank $R$ is denoted by $N_R$
and it is also indicated in Table 2. We define the fraction of 
elements having the cultural vector with rank $R$ as $C_R=N_R/N$.
Note that about half of the elements in the final state share 
its cultural vector among the seven most common vectors. 

Figure 1 shows the pattern of the final state of the system.
The labels indicate 
the rank of the cultural vector corresponding to each site. 
Contiguous sites having identical cultural vectors are joined by lines.
Elements with identical cultural vectors tend to form domains, in spite 
that neighbors with overlap $F-1$ do not interact.

The number of surviving traits monotonically decreases during the evolution of 
the system toward its final state.
When $F_d=0$ 
the number of surviving traits in the final state is
$F$ if $q<q_c$ (all elements are identical and
therefore there is one trait per feature). As shown above, for $F_d>0$ the
number of surviving traits in the final state is greater than $F$ but
much smaller than its maximum possible value of $F \times q$.
On the other hand, the size of the largest domain is equal to $N$ for
$F_d=0$ and $q<q_c$, while for $F_d>0$ the largest group
involves only a fraction of the elements in the system.
Figure~2 shows the evolution of both, the number of surviving traits and
the fraction of elements having the most abundant cultural vector,
$C_1=N_1/N$, in a system of size $N=120 \times 120$, 
for $F_d=0$ and $F_d=1$.

\section{Statistical properties}
Figure 3 shows the average fraction of elements with the most abundant vector 
$\langle C_1 \rangle$ as a function of $q$, for several values of the parameter $F_d$ in a 
system of size $N=40 \times 40$ and $F=11$. For $F_d=0$, there exist a
critical value $q_c$ at which the order parameter $\langle C_1 \rangle$ exhibits an
abrupt transition from a homogeneous, monocultural state, characterized 
by $\langle C_1 \rangle=1$, to a disordered, multicultural state, 
where $\langle C_1 \rangle \ll 1$ \cite{Castellano00}.
The value $q_c$ is not very sensitive to the variation of the parameter $F_d$.
For $F_d > 0$ and $q < q_c$, the value of $\langle C_1 \rangle$ is less than $1$, 
indicating the presence of 
intracultural diversity in the single surviving cultural zone (there are no
neighbors with zero overlap).

Figure 4 shows 
the average fraction of 
cultural vectors $\langle C_R \rangle$ by rank $R$, 
for different system sizes.
Each curve is the average over $50$ realizations of initial conditions.
For the parameter values used in Fig. 4 ($q \ll q_c$) there is a
single cultural zone in the final state but 
several cultural vectors survive having a distribution that 
depends on the system size.
The dispersion on each curve is   
larger for low and for high values of $R$ than for intermediate values of the
rank. The dispersion for low values of $R$ reflects the competition
between the more abundant vectors during the evolution towards the 
frozen state. On the other hand, the dispersion for large values of $R$
is mostly due to the fluctuations on the total number of different vectors
present in the final state.  

The frozen patterns in a cultural zone are complex, as shown in
Figure 1. The cultural diversity can be characterized 
in terms of the distribution of the cultural affinity between any two
elements $i$ and $j$ in the system, defined as
$A(i,j) = \omega(i,j)/F$.
Figure 5 shows the average distribution of the
cultural affinities $A(i,j)$ of all pairs of elements in
the system, averaged over $10$ realizations. 
The three curves on each panel in Fig. 5 correspond to
three different values of $q$.
In the case that $F_d=0$, the distribution for $q < q_c$ consists of a single
peak at the value $A(i,j)=1$, corresponding a homogeneous state,
while for $q > q_c$ a second peak appears
at $A(i,j)=0$, reflecting the presence of multiple domains. 
As shown in Fig. $5$, for $F_d>0$ intracultural
diversity is manifested as a wide spectrum of 
cultural affinities present in the system.
As $F_d$ increases, the distribution of cultural affinities shifts
towards $A(i,j)=0$, reflecting the increase of intracultural
diversity within the cultural zones. 

The spatial distribution of cultural diversity can be characterized
by the average shell affinity $S(r)$ defined as 
\begin{equation}
S(r)= \frac{1}{N \times n(r)} \, 
\sum_{i=1}^N \sum_{j \in \rho(i,r)} A(i,j),
\end{equation}
where $\rho(i,r)$ is the set of elements in a square shell of
radius $r$ centered in element $i$
(the unit of distance is one site),
and $n(r)$ is the number of elements on this shell.

Figure 6 shows $S(r)$ for several values of $F_d$.
The average shell affinity $S(r)$ is well fitted by the relation
\begin{equation}
S(r) \simeq S(1) - \alpha \log r.
\end{equation}
We find the scaling $S(1) \simeq 1-F_d/F$ and 
$\alpha \simeq 1.43 F_d/F$
for a wide range of values of $F$, $F_d$ and $q<q_c$.
The slope $\alpha= - \frac{d{S(r)}}{d{\log(r)}}$ characterizes the
gradient of intracultural diversity or {\em cultural cline}.
Extrapolation of equation $(2)$ to $S(r_0)=0$ allows for a definition of 
a characteristic distance
$r_0=\exp \left[ 0.7 (\frac{F}{F_d}-1) \right]$  
between two elements that have zero overlap; a concept that can be
applied to the study of the spatial distribution of related cultures. 
These results suggest that the relevant quantity for the description of
intracultural diversity is the ratio $F_d/F$.

\section{Conclusions}
In order to allow for individual 
nonconformity in the Axelrod's model
for cultural dissemination, we have introduced a parameter $F_d$
that reduces the maximum number
of shared features for interaction.
The inclusion of this parameter maintains the main features
of the Axelrod's model, corresponding to $F_d=0$.
In particular, there is a nonequilibrium transition from a single cultural zone to
a multicultural state at about the same critical value $q_c$ at which
the Axelrod's model exhibits a transition from a homogeneous,
monocultural region, to a multicultural state.
However, the addition of parameter $F_d$ 
set the stage for the occurrence of
complex patterns of intracultural diversity in cultural zones.
The intracultural diversity associated to the constraint 
$F_d > 0$ is manifested in the appearance of a distribution
of the abundance of cultural vectors by rank
inside cultural zones. We found that this distribution is sensitive to the size of 
the zone, as shown in Figure 4.

We have introduced the concept of cultural affinity between
any two elements in order to
characterize intracultural diversity in the system. 
The cultural affinity among all the elements in the system for $F_d>0$
shows a wide distribution in contrast to the case $F_d=0$ where the
cultural affinity can only take the values $1$ when $q<q_c$, or $1$ and $0$
when $q>q_c$.

The cultural affinity between elements separated
by a given distance $S(r)$ has led us to the concept
of cultural cline defined as 
$\alpha= - \frac{d{S(r)}}{d{\log(r)}}$.
We found that $S(r)$ depends only on the ratio $F_d/F$ since
it is well fitted by the relation
$S(r) \simeq 1-F_d/F - 1.43 F_d/F \log r$
for a wide range of values of $F$, $F_d$ and $q<q_c$.

The model presented here may be useful to describe the emergence
of cultural gradients such as dialects, gastronomic customs, etc, 
in geographical areas. 
In the biological context, this model can also be 
adapted to the study of phenotype clines. For instance, it have been proposed that 
the evolution of female mating preferences can greatly amplify large-scale 
geographic variation in male secondary sexual characters and produce 
widespread reproductive isolation with no geographic 
discontinuity \cite{Russell82}.

\section*{Acknowledgments}
This work was supported by Consejo de Desarrollo Cient\'ifico,
Human\'istico y Tecnol\'ogico of the Universidad de Los Andes,
M\'erida,  under grant No. C-1275-05-05-B  and by FONACIT,
Venezuela, under grant No. F-2002000426. M.G.C. thanks the 
Condensed Matter and Statistical Physics Section of the Abdus Salam
International Centre for Theoretical Physics for support and hospitality
while part of this work was carried out.

\eject
\begin{tabular}{c}
\hspace{3.7cm}TABLE 1\\
\hspace{3.7cm}$K_{f}(s)$: Traits present in the final state\\
\end{tabular}\\
\begin{tabular}{ccccccccccccl}
\hline
\hline
& & \vline &  & & & & & \hspace{-1.7cm} Trait ($s$)&  &  &  & \\
& & \vline & 0 &1 &2 &3 & 4 & 5 & 6 & 7 & 8 & 9\\
&Feature ($f$) &\vline &---- &---- &---- &---- &---- &---- &---- &---- &---- &---- \\
& 1&\vline &   0& 400&   0&   0&   0&   0&   0&   0&   0&   0\\
& 2&\vline &   0& 280&  63&  57&   0&   0&   0&   0&   0&   0\\
& 3&\vline &   0&   0&   0&   0&   0& 361&   0&  39&   0&   0\\
& 4&\vline &   0&   0& 225&   0&   0& 175&   0&   0&   0&   0\\
& 5&\vline &  49&   0&   0&   0&  36&   0&   0&   0& 315&   0\\
& 6&\vline &   0&   0&   0& 400&   0&   0&   0&   0&   0&   0\\
& 7&\vline & 292& 108&   0&   0&   0&   0&   0&   0&   0&   0\\
& 8&\vline &   0& 132&   0&   0&   0& 268&   0&   0&   0&   0\\
& 9&\vline &   0&   0&   0&   0&   0&   0&   0& 400&   0&   0\\
&10&\vline &   0&   0&   0&   0&   0& 400&   0&   0&   0&   0\\
&11&\vline &   0&   0&  27&   0&   0& 373&   0&   0&   0&   0\\
\hline
\hline
\end{tabular}

\eject

\begin{tabular}{c}
\hspace{3.7cm}TABLE 2\\
\hspace{3.7cm}More common vectors in the final state\\
\end{tabular}\\
\begin{tabular}{ccccccccccccccl}
\hline
\hline
rank& occurrence&accumulated&\vline &&&&&&&&\hspace{-3.2cm}VECTOR&&&\vspace{-0.35cm}\\
    &          &fraction&\vline &&&&&&&&&&&\\
\hline
 1& 58& 0.145&\vline & (1,& 1,& 5,& 2,& 8,& 3,& 0,& 5,& 7,& 5,& 5) \\
 2& 52& 0.275&\vline & (1,& 1,& 5,& 2,& 8,& 3,& 0,& 1,& 7,& 5,& 5) \\
 3& 28& 0.345&\vline & (1,& 1,& 5,& 5,& 8,& 3,& 0,& 5,& 7,& 5,& 5) \\
 4& 23& 0.403&\vline & (1,& 1,& 5,& 5,& 8,& 3,& 1,& 5,& 7,& 5,& 5) \\
 5& 16& 0.443&\vline & (1,& 2,& 5,& 5,& 8,& 3,& 0,& 5,& 7,& 5,& 5) \\
 6& 14& 0.478&\vline & (1,& 1,& 5,& 5,& 0,& 3,& 1,& 5,& 7,& 5,& 5) \\
 7& 13& 0.510&\vline & (1,& 1,& 5,& 2,& 8,& 3,& 1,& 5,& 7,& 5,& 5) \\
 8& 13& 0.542&\vline & (1,& 1,& 5,& 2,& 8,& 3,& 0,& 1,& 7,& 5,& 2) \\
 9& 12& 0.573&\vline & (1,& 1,& 5,& 2,& 4,& 3,& 0,& 1,& 7,& 5,& 5) \\
10& 11& 0.600&\vline & (1,& 2,& 5,& 5,& 8,& 3,& 1,& 5,& 7,& 5,& 5) \\
11& 11& 0.627&\vline & (1,& 3,& 5,& 2,& 8,& 3,& 0,& 1,& 7,& 5,& 5) \\
12& 10& 0.652&\vline & (1,& 1,& 7,& 2,& 8,& 3,& 0,& 5,& 7,& 5,& 5) \\
13&  9& 0.675&\vline & (1,& 3,& 5,& 2,& 4,& 3,& 0,& 1,& 7,& 5,& 5) \\
14&  8& 0.695&\vline & (1,& 2,& 7,& 5,& 8,& 3,& 1,& 5,& 7,& 5,& 5) \\
15&  8& 0.715&\vline & (1,& 2,& 7,& 5,& 8,& 3,& 0,& 5,& 7,& 5,& 5) \\
\hline
\hline
\end{tabular}

\eject

\begin{figure}
\includegraphics[width=14cm]{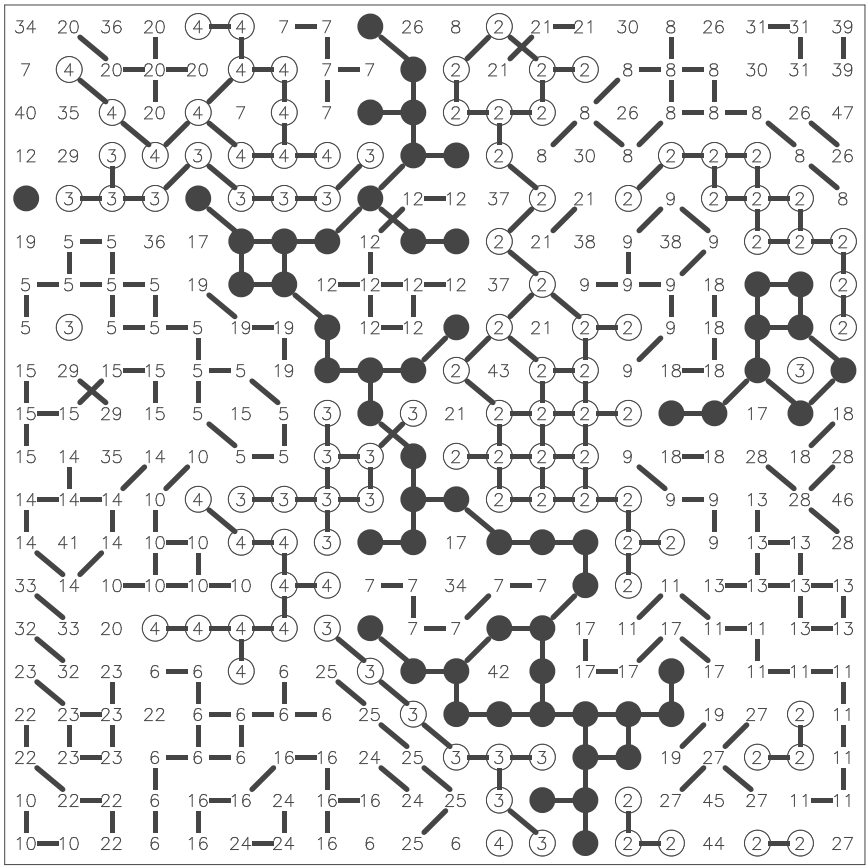}
\caption{ Spatial pattern corresponding to a final state for a system of
size $N=20 \times 20$ with parameters $F=11$, $F_d=1$, and $q=10$.
The numbers on each site indicate the rank of the cultural vector of that site.
The most common vectors (rank $R=1$) are plotted as solid circles. 
Sites whose vectors have ranks $R=2,3,4$ are plotted as open circles.
In order to facilitate the recognition of patterns
a line between contiguous sites having identical states are connected by lines.
}
\end{figure}

\begin{figure}
\includegraphics[width=14cm]{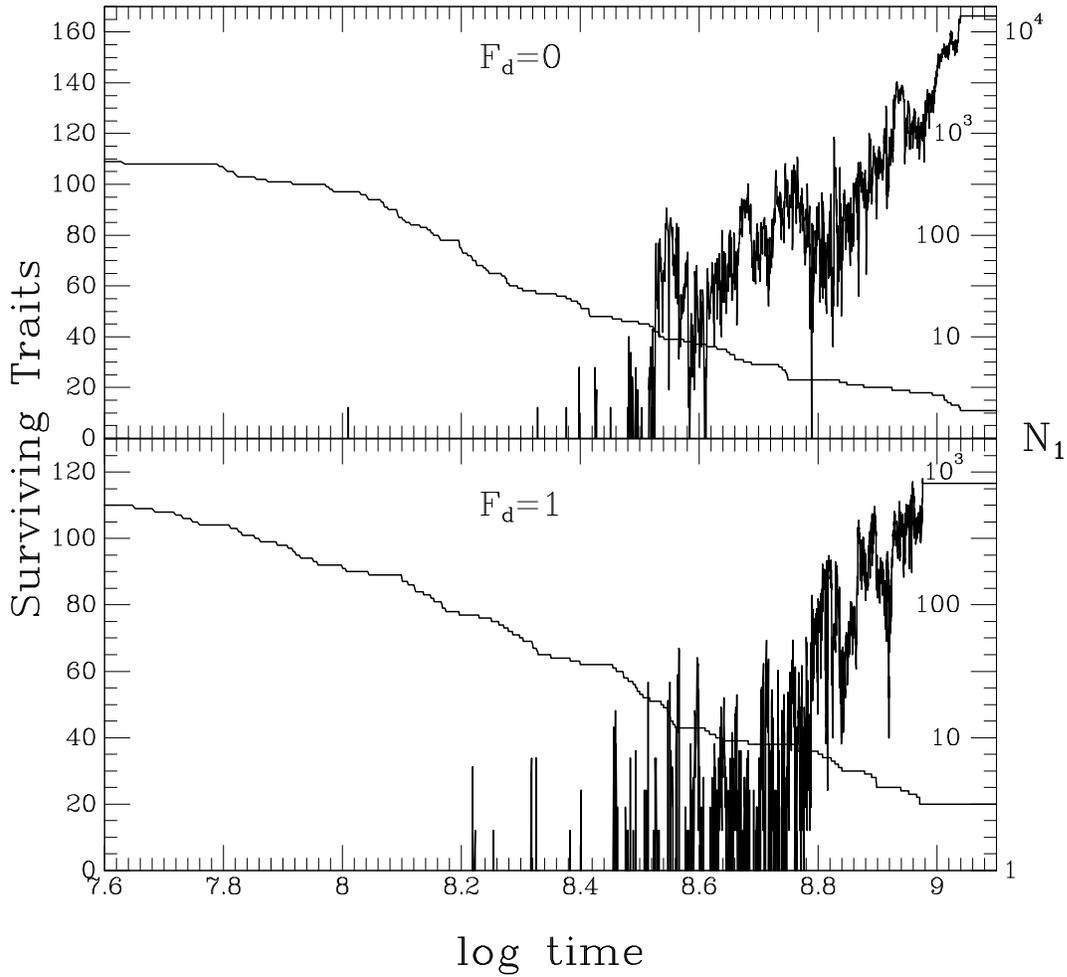}
\caption{
Evolution for a system of
size $N=120 \times 120$ with parameters $F=11$ and $q=10$,
for $F_d=0$ (upper panel) and $F_d=1$ (lower panel). 
The monotonically decreasing curve on each panel corresponds
to the number of surviving traits (left vertical axis) as a function of 
time. 
The fluctuating curve on each panel shows the size of
the largest cultural group $N_1$ (right vertical axis) as a function of
time.
Time is measured as the number of iterations that 
result in a trait exchange.
}
\end{figure}

\begin{figure}
\includegraphics[width=14cm]{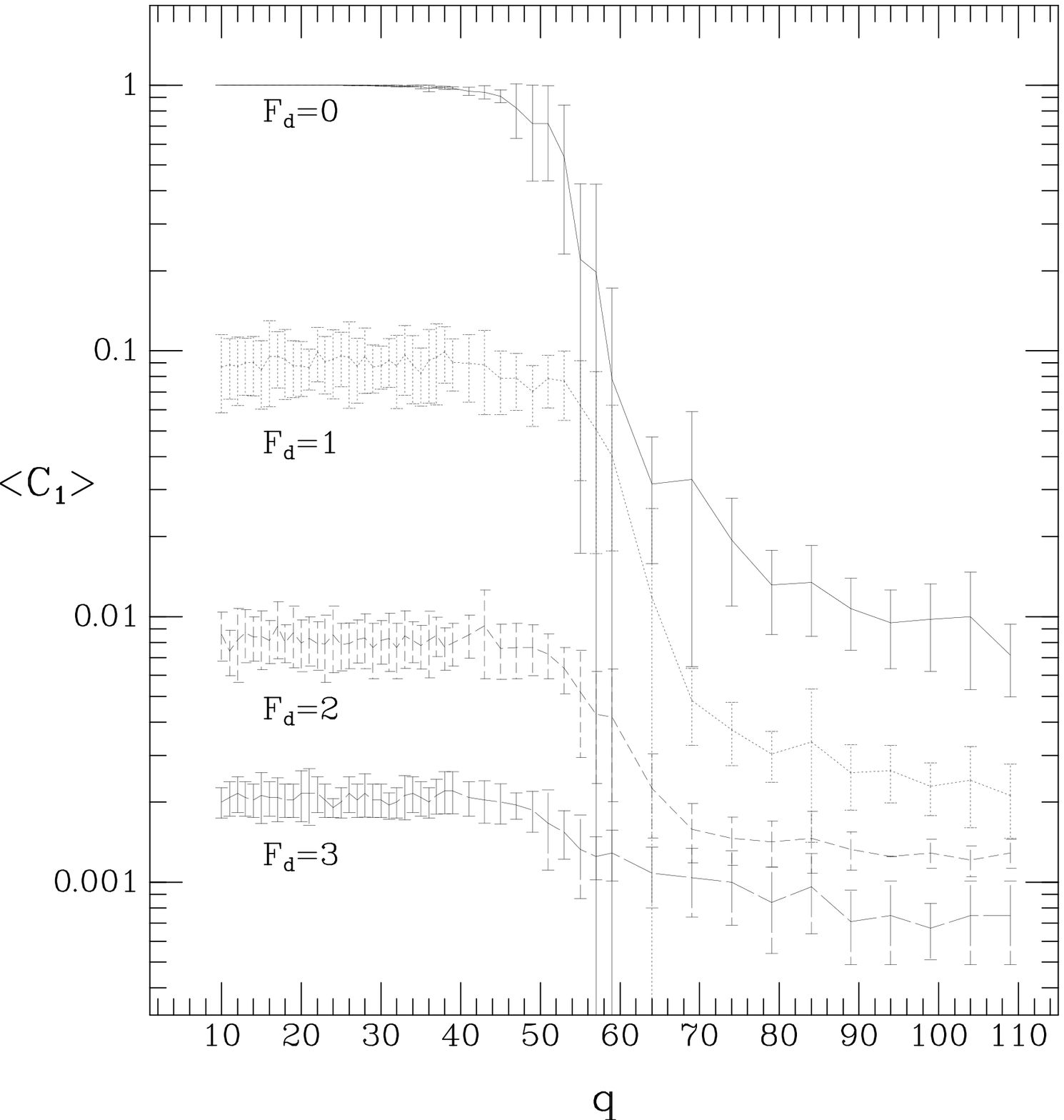}
\caption{
Average fraction of elements having the most abundant cultural vector,
$\langle C_1 \rangle$, as a function of $q$, for several values of $F_d$ in a
system of size $N=40 \times 40$ and $F=11$.
The label on each curve indicates the value of $F_d$.
Error bars corresponding to one standard deviation are shown on each curve.
}
\end{figure}

\begin{figure}
\includegraphics[width=14cm]{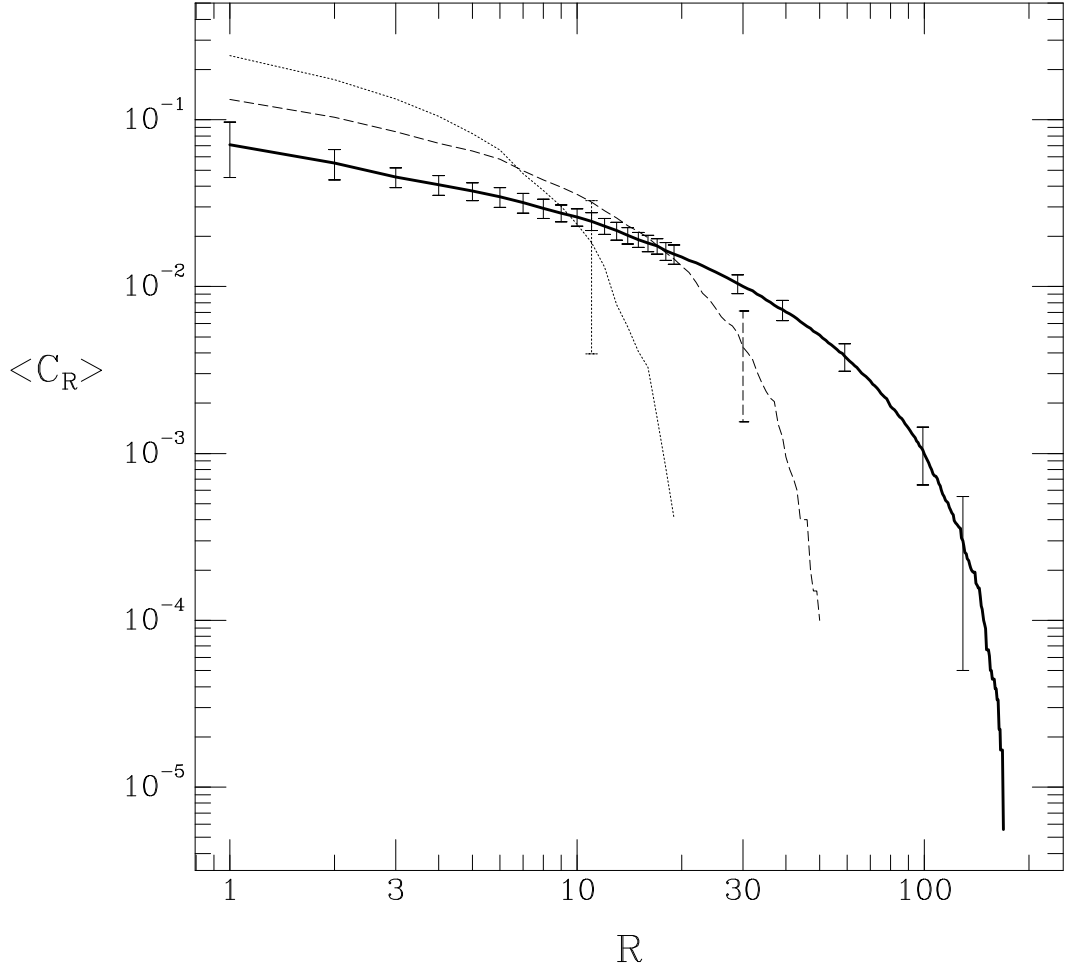}
\caption{
Average fraction of
cultural vectors $\langle C_R \rangle$ as a function of the rank $R$,
for different system sizes with fixed parameter values 
$F_d=1$, $F=11$ and $q=10 < q_c$.
Continuous, dashed and dotted lines correspond to $N=60 \times 60$, 
$20 \times 20$ and $7 \times 7$, respectively.  
Error bars corresponding to one standard deviation are shown
on several points.
}
\end{figure}

\begin{figure}
\includegraphics[width=14cm,angle=0]{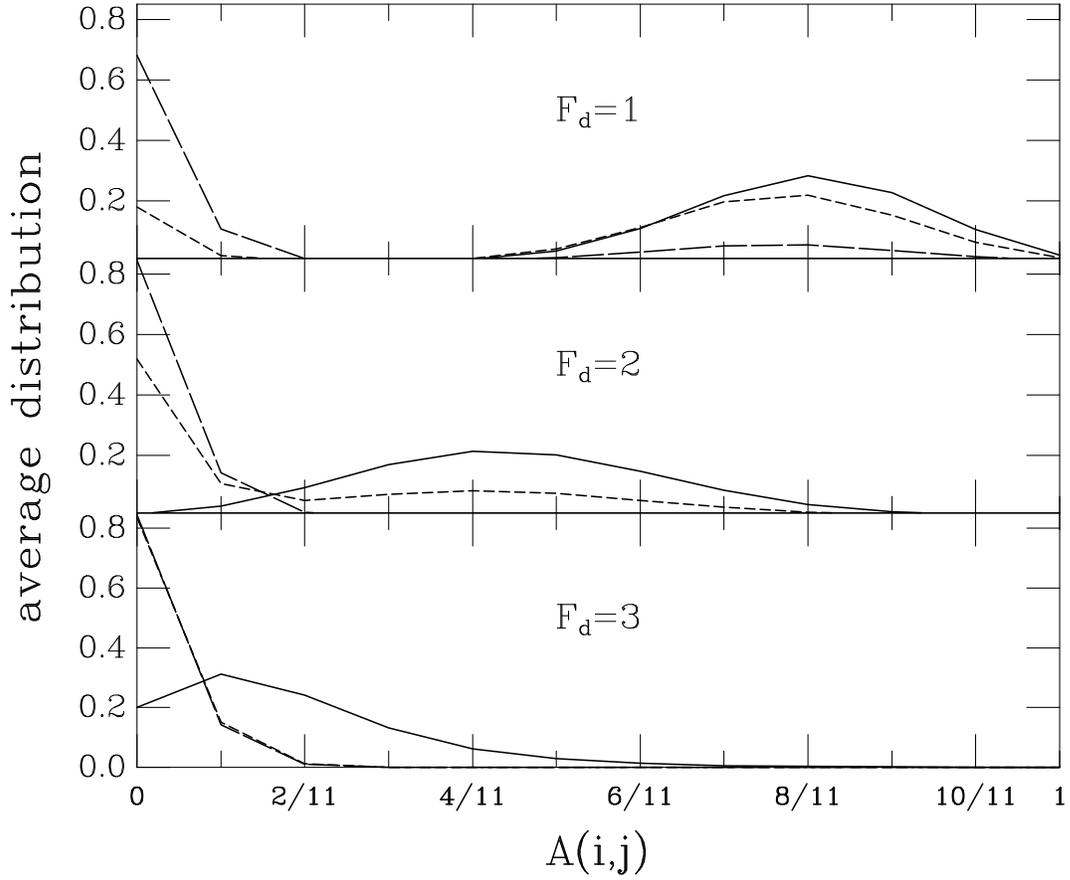}
\caption{
Average distribution of the
cultural affinities $A(i,j)$ of all pairs of elements in
the system, averaged over $10$ realizations, for the value of $F_d$
indicated on each panel.
The continuous, the dashed, and the long-dashed lines 
on each panel correspond to the values of $q=10$, $q=62$, 
and $q=66$, respectively.
System size is $N=80 \times 80$, and $F=11$.
}
\end{figure}

\begin{figure}
\includegraphics[width=14cm,angle=0]{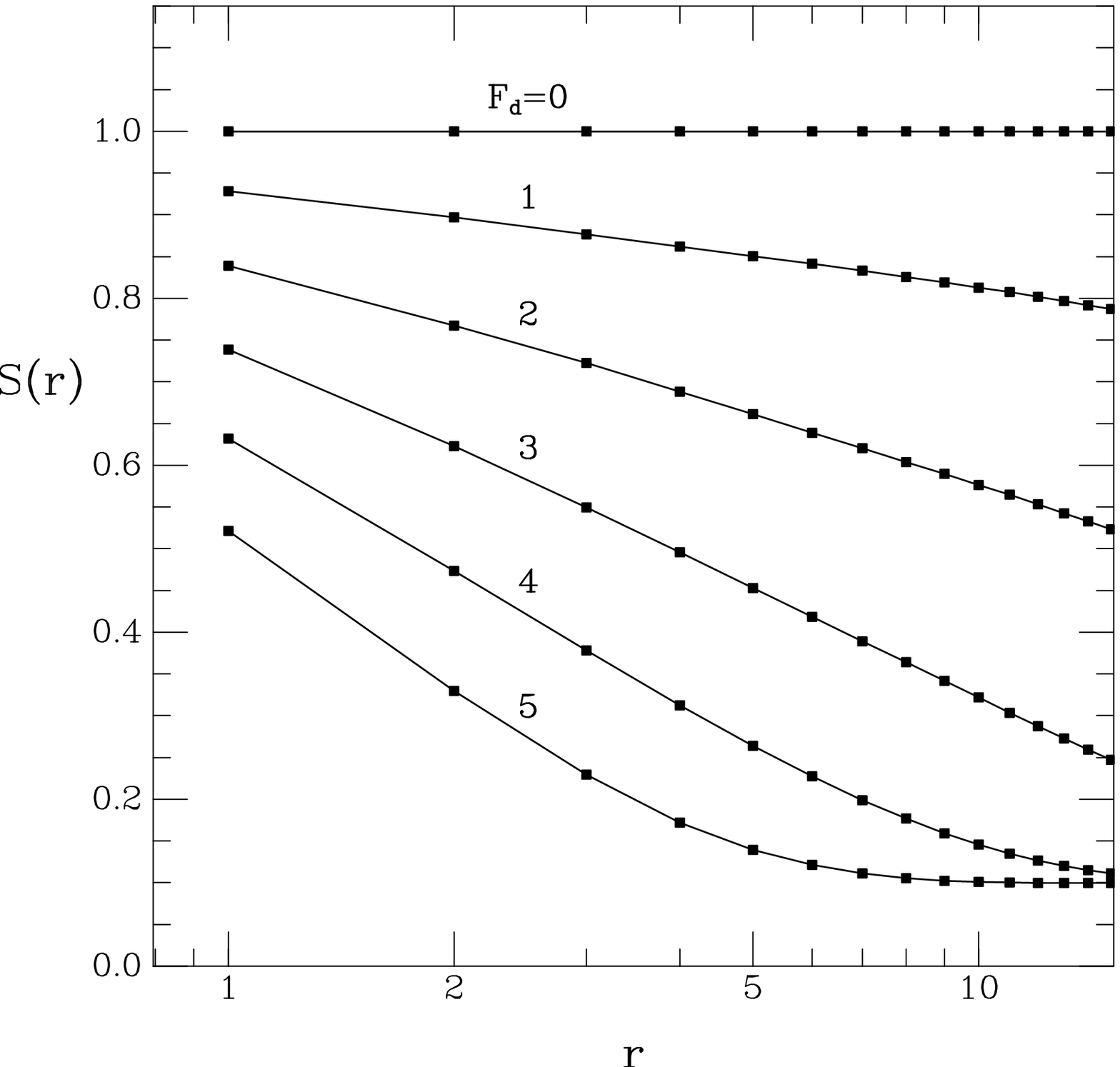}
\caption{
The average shell affinity $S(r)$ for different values of $F_d$
as indicated by the label besides each curve.
System size is $N=80 \times 80$, $F=11$, and $q=10$.
The slope of each curve, $\alpha= - \frac{d{S(r)}}{d{\log(r)}}$, 
characterizes the {\em cultural cline}.
}
\end{figure}


\begin{thebibliography}{99}

\bibitem{Weidlich}
        W. Weidlich, Phys. Rep. {\bf 204} (1991) 1.
\bibitem{Stauffer}
        S. M. de Oliveira, P. M. C. de Oliveira, and D. Stauffer,
        {\it Nontraditional Applications of Computational Statistical Physics}
        (B. G. Teubner, Stuttgart, 1999).
\bibitem{Arrow}
        P.W. Anderson, K. Arrow, and D. Pines, {\it The Economy as
        an Evolving Complex System} (Addison-Wesley, Redwood, 1998).
\bibitem{MaxiR}
         M. San Miguel, V. Eguiluz, R. Toral, and K. Klemm,
         Computing in Science and Engineering {\bf 7} (2005) 67.
\bibitem{axel97}
        R. Axelrod, J. Conflict Res. {\bf 41} (1997) 203.
\bibitem{Castellano00}
         C. Castellano, M. Marsili, A Vespignani, Phys. Rev. Lett. {\bf85} (2000)
         3536.
\bibitem{Vilone}
         D. Vilone, A. Vespignani, and C. Castellano,
         Eur. Phys. J. B {\bf 30} (2002) 299.
\bibitem{Klemm03a}
         K. Klemm, V. M. Eguiluz, R. Toral, and M. San Miguel,
         Phys. Rev. E {\bf 67} (2003) 026120.
\bibitem{Maxi3} 
         K. Klemm, V. Eguiluz, R. Toral, M. San Miguel,
         Physica A, {\bf 327} (2003) 1.
\bibitem{Klemm03b}
         K. Klemm, V. M. Eguiluz, R. Toral, and M. San Miguel,
         Phys. Rev. E  {\bf 67} (2003) 045101(R).
\bibitem{GCT}
         J. C. Gonz\'alez-Avella, M. G. Cosenza, and K. Tucci, 
         Phys. Rev. E {\bf 72} (2005) 065102(R).
\bibitem{Shibanai}
         Y. Shibanai, S. Yasuno, and I. Ishiguro,
         J. Conflict Res.  {\bf 45} (2001)  80.
\bibitem{PRE}
         J. C. Gonz\'alez-Avella, V. M. Egu\'iluz, M. G.
         Cosenza, K. Klemm, J.L. Herrera and M. San Miguel, Phys. Rev. E
         {\bf 73} (2006) 046119.
\bibitem{Greig}
         J. Greig, J. of Conflict Resolution {\bf 46} (2002) 225.
\bibitem{Kuperman}
         M. N. Kuperman, Phys. Rev. E  {\bf 73} (2006) 046139.
\bibitem{Flache1}
         A. Flache and M. Macy, Los Alamos Arxiv physics/0604201 (2006).
\bibitem{Russell82}
         R. Lande, Evolution {\bf 36} (1982) 213.

\end{thebibliography}
\end{document}